# Gaussian, Hermite-Gaussian, and Laguerre-Gaussian beams: A primer

Francesco Pampaloni,[1] Jörg Enderlein[2]

[1]EMBL, Cell Biology and Biophysics Programme, Meyerhofstr. 1

D-69117 Heidelberg, Germany

[2]Institute for Biological Information Processing 1

Forschungszentrum Jülich, D-52425 Jülich, Germany

**Abstract**

The paper aims at presenting a didactic and self-contained overview of Gauss-Hermite and Gauss-Laguerre laser beam modes. The usual textbook approach for deriving these modes is to solve the Helmoltz electromagnetic wave equation within the paraxial approximation. Here, a different technique is presented: Using the plane wave representation of the fundamental Gaussian mode as seed function, all higher-order beam modes are derived by acting with differential operators on this fundamental solution. Even special beam modes as the recently introduced Bessel beams are easily described within that framework.

**Introduction**

The description of the spatial structure of the different beam modes has been object of theoretical and experimental analysis since the invention of the laser, and now is a standard textbook topic in modern optics and laser physics courses. The usual starting point for the derivation of laser beam propagation modes is solving the scalar Helmholtz equation within the paraxial approximation [1-3]. In this paper we present a didactic and self-contained overview of the different laser beam modes by introducing a unified theoretical treatment based on a plane wave representations of the electromagnetic field. This approach can be easily generalized to non-paraxial situations and to include vector effects of the electromagnetic field. All higher-order beams, like for instance Hermite-Gaussian, Laguerre-Gaussian, and Bessel beams can elegantly be derived by acting with differential operators on the plane wave representation of the fundamental Gaussian mode used as a seed function [4-9]. Still recently, higher-order laser beams were the object of study of a restricted group of specialists. Thank to several new applications especially in the field of optical trapping, the interest concerning these beams is now rising rapidly. Optical tweezers with more sophisticated features can be realized with higher-order beams. Laguerre-Gaussian beam can transfer orbital momentum to a trapped particle, inducing it to rotate around the optical axis. For this reason, Laguerre-Gaussian optical tweezers are also known as "optical vortices" [10-13]. Laguerre-Gaussian tweezers can also trap metallic particles or particles with a refractive index higher than that of the surrounding medium [14, 15]. So-called Bessel beams [16, 17] are a further interesting example of higher-order beams which can be conveniently described by the plane wave representation. Bessel beams are also known as "non-diffracting" modes, because they have the property of self-reconstructing after encountering an obstacle [18]. Several new promising applications of Bessel beams are emerging. For example, simultaneous trapping of several particles over large axial distances (in the range of millimeters) is possible by using just one single Bessel beam [19].

This paper is organized as follows. First, the Gaussian fundamental mode is introduced as the beam solution jointly minimizing both beam divergence and beam diameter. Basic concepts such as Rayleigh length, beam waist, wave front curvature, and Gouy's phase are introduced. Following, Hermite-Gauss beams with *complex* arguments in the Hermite functions (also called "elegant" Hermite-Gaussian modes in the literature [1, 5, 6]) are derived by differentiating the fundamental Gauss mode with respect to the transverse coordinates. Next, it is shown how this derivation can be generalized to obtain a whole family of Hermite-Gauss modes, including also those with *real* arguments in the Hermite functions (termed also "standard" Hermite-Gaussian modes). The same approach is subsequently extended to derive Laguerre-Gaussian modes. Finally, Bessel beams are presented as special cases of plane wave representations, and the concepts of non-diffractivity and self-reconstruction are briefly discussed.

**The fundamental Gaussian mode**

In the scalar field approximation (e.g. neglecting the vector character of the electromagnetic field), any electric field amplitude distribution can be represented as a superposition of plane waves, i.e. by

$$E \propto \iint\limits_{k_x^2+k_y^2 \leq k^2} \frac{dk_x dk_y}{(2\pi)^2} A(k_x, k_y) \exp\left(ik_x x + ik_y y + i\sqrt{k^2 - k_x^2 - k_y^2}\, z\right) \qquad (1)$$

This representation is also called angular spectrum of plane waves or plane-wave expansion of the electromagnetic field [7-9]. In (1), $A(k_x, k_y)$ is the amplitude of the plane wave contribution with wave vector $\mathbf{k} = \left(k_x, k_y, \sqrt{k^2 - k_x^2 - k_y^2}\right)$ (in Cartesian coordinates). The modulus of the wave vector, $k = |\mathbf{k}|$, is equal to $2\pi n/\lambda$, where $n$ is the refractive index of the medium, and $\lambda$ the wavelength of light in vacuum. This representation automatically obeys the scalar wave equation

$$\frac{\partial^2 E}{\partial x^2} + \frac{\partial^2 E}{\partial y^2} + \frac{\partial^2 E}{\partial z^2} + k^2 E = 0, \qquad (2)$$

for an arbitrary choice of the amplitude function $A(k_x, k_y)$, which can be directly demonstrated by substituting Eq.(1) into Eq.(2). Representation (1) is chosen in such a way that the net energy flux connected with the electromagnetic field is towards the propagation axis $z$. Every plane wave is connected with an energy flow that has direction $\mathbf{k}$ and value $\frac{nc}{8\pi}|A(k_x, k_y)|^2$, $n$ being the refractive index of the medium, and $c$ denoting the speed of light in vacuum.

Actual lasers generate a spatially coherent electromagnetic field which has a finite transversal extension and propagates with moderate spreading. That means that the wave amplitude changes only slowly along the propagation axis ($z$-axis) compared to the wavelength and finite width of the beam. Thus, the paraxial approximation can be applied to Eq.(1), assuming that the amplitude function $A(k_x, k_y)$ falls off sufficiently fast with increasing values of $k_{x,y}$ so that the integration limits can be extended to plus/minus infinity, and that the root occurring in the exponent in Eq.(1) can be expanded into $k - (k_x^2 + k_y^2)/2k$. Following this assumption, only values with $k_{x,y} \ll k$ significantly contribute to the integral's final value.

Two principal characteristics of the total energy flux can be considered: the divergence (spread of the plane wave amplitudes in wave vector space), defined as

$$\text{divergence} \propto \int_{-\infty}^{\infty} \frac{dk_x}{2\pi} \int_{-\infty}^{\infty} \frac{dk_y}{2\pi} \left(k_x^2 + k_y^2\right)|A|^2$$

and the transversal spatial extension (spread of the field intensity perpendicular to the *z*-direction) defined as

$$\text{transversal extension} \propto \int_{-\infty}^{\infty} dx \int_{-\infty}^{\infty} dy \left(x^2 + y^2\right)|E|^2 = \int_{-\infty}^{\infty} \frac{dk_x}{2\pi} \int_{-\infty}^{\infty} \frac{dk_y}{2\pi} \left(\left|\frac{\partial A}{\partial x}\right|^2 + \left|\frac{\partial A}{\partial y}\right|^2\right).$$

The transition from the l.h.s. to the r.h.s. in the last expression is obtained by inserting representation Eq.(1) for *E* and carrying out the integrations over *x* and *y*.

Let us seek the fundamental mode of a laser beam as the electromagnetic field having simultaneously minimal divergence and minimal transversal extension, i.e. as the field that minimizes the product of divergence and extension.

By symmetry reasons, this leads to looking for an amplitude function minimizing the product

$$\left(\int_{-\infty}^{\infty} \frac{dk_x}{2\pi} k_x^2 |A|^2\right)\left(\int_{-\infty}^{\infty} \frac{dk_x}{2\pi} \left|\frac{\partial A}{\partial k_x}\right|^2\right) \tag{3}$$

and a similar product for the *y*-direction. Next, one can employ the Schwartz inequality which states that, for two arbitrary complex functions $f(\xi)$ and $g(\xi)$,

$$\int_{-\infty}^{\infty} d\xi |f(\xi)|^2 \int_{-\infty}^{\infty} d\xi |g(\xi)|^2 \geq \left|\int_{-\infty}^{\infty} d\xi \overline{f(\xi)} g(\xi)\right|^2, \tag{4}$$

provided that all integrals converge, and where a bar denotes complex conjugation. Note that the equal sign in this relation holds *only* if *f* and *g* differ solely by a constant factor. Applying Schwartz' inequality to the product of Eq.(3) one obtains

$$\left(\int_{-\infty}^{\infty}\frac{dk_x}{2\pi}k_x^2|A|^2\right)\left(\int_{-\infty}^{\infty}\frac{dk_x}{2\pi}\left|\frac{\partial A}{\partial k_x}\right|^2\right) \geq \left|\int_{-\infty}^{\infty}\frac{dk_x}{(2\pi)^2}\frac{1}{2}\left(k_x\overline{A}\frac{\partial A}{\partial k_x}+k_xA\frac{\partial \overline{A}}{\partial k_x}\right)\right|^2$$

$$= \left|\int_{-\infty}^{\infty}\frac{dk_x}{(2\pi)^2}\frac{k_x}{2}\frac{\partial |A|^2}{\partial k_x}\right|^2 \qquad (5)$$

$$= \left|\frac{1}{8\pi^2}\int_{-\infty}^{\infty}dk_x|A|^2\right|^2 = \frac{\|A\|^4}{(8\pi^2)^2}$$

where a partial integration was used in the transition from the middle to the last line, assuming that the amplitude $A$ vanishes for $k_x \to \pm\infty$. Thus, if the norm $\|A\|$ is kept constant, the product of divergence and extension reaches its minimum when the l.h.s. in the last equation becomes equal to the r.h.s., which is only the case if $k_x A$ and $\partial A/\partial k_x$ differ solely by a constant factor (property of the Schwartz inequality, see above). A similar result holds for the $y$-direction, which leads to the explicit amplitude expression

$$A(k_x,k_y) \propto \exp\left(-\frac{w_0^2}{4}(k_x^2+k_y^2)\right) \qquad (6)$$

which contains the arbitrary proportionality factor $-w_0^2/2$. Returning to the electric field amplitude, the fundamental laser mode is then written as

$$E_0 \propto \int_{-\infty}^{\infty}\frac{dk_x}{2\pi}\int_{-\infty}^{\infty}\frac{dk_y}{2\pi}\exp\left(ik_xx+ik_yy+ikz-i\frac{k_x^2+k_y^2}{2k}z-\frac{w_0^2}{4}(k_x^2+k_y^2)\right). \qquad (7)$$

The integration in Eq.(7) can be carried out by employing the general relation

$$\int_{-\infty}^{\infty}\frac{d\xi}{2\pi}\exp\left(i\xi x-\frac{\sigma^2}{2}\xi^2\right) = \frac{1}{\sqrt{2\pi\sigma^2}}\exp\left(-\frac{x^2}{2\sigma^2}\right) \qquad (8)$$

leading to

$$E_0 = \frac{1}{w_0^2 + 2iz/k} \exp\left[ikz - \frac{x^2 + y^2}{w_0^2 + 2iz/k}\right], \tag{9}$$

where a constant factor π was omitted. Introducing the Rayleigh length $z_R$, (which is the distance from the beam focus at which the beam area doubles)

$$z_R = \frac{kw_0^2}{2} = \frac{\pi w_0^2}{\lambda} \tag{10}$$

(λ being the vacuum wavelength) and the reduced coordinate ζ

$$\zeta = z/z_R \tag{11}$$

and employing the fact that

$$\frac{1}{1+i\zeta} = \frac{1-i\zeta}{1+\zeta^2} = \frac{\exp(-i\arctan\zeta)}{\sqrt{1+\zeta^2}} \tag{12}$$

the final result for the electric field amplitude is found as

$$E_0 = \frac{1}{w_0\sqrt{1+\zeta^2}} \exp\left[ikz - \frac{1-i\zeta}{w_0^2(1+\zeta^2)}\rho^2 - i\arctan(\zeta)\right] \tag{13}$$

where another constant $w_0$ was omitted and the abbreviation $\rho^2 = x^2 + y^2$ was used.

The field mode as given by Eq.(13) is called the *fundamental Gaussian mode* because the corresponding light intensity

$$|E_0|^2 = \frac{1}{w_0^2(1+\zeta^2)} \exp\left[-\frac{2\rho^2}{w_0^2(1+\zeta^2)}\right] \tag{14}$$

shows a Gaussian profile perpendicularly to the propagation axis $z$.

Thus, seeking the field with minimal divergence and minimal transversal extension has led directly to the fundamental Gaussian beam. This means that the Gaussian beam is the mode with minimum uncertainty, i.e. the product of its sizes in real space and wave-vector space is the theoretical minimum as given by the Heisenberg's uncertainty principle of Quantum Mechanics. Consequently, the Gaussian mode has less dispersion than any other optical field of the same size, and its diffraction sets a lower threshold for the diffraction of real optical beams. The diameter of the Gaussian beam is defined by

$$w(\zeta) = w_0 \left(1 + \zeta^2\right)^{1/2}. \tag{15}$$

defining the radius where the electric field intensity $|E_0|^2$ has fallen off to $1/2e^2$ of its maximum value at $\rho = 0$. The $\arctan(\zeta)$-term in Eq.(13) is called Gouy's phase $\psi_0(\zeta)$,

$$\psi_0(\zeta) = \arctan \zeta, \tag{16}$$

Which describes the rapid phase change of the electric field when traversing the point of minimal beam diameter at $\zeta = 0$ (beam waist). Finally, the term $i\zeta\rho^2/w^2(\zeta)$ in Eq.(13) has the consequence that the surfaces of constant phase are not planes perpendicular to the axis of propagation, $z$, but are curved surfaces. The curvature of these surfaces can be found by looking at the condition of stationary phase

$$kz + \frac{\zeta\rho^2}{w^2(\zeta)} - \psi_0 \approx kz + \frac{\zeta\rho^2}{w^2(\zeta)} = \text{const.}, \tag{17}$$

defining parabolas $z = -\zeta\rho^2/kw^2(\zeta)$ with apex curvature radius

$$R(\zeta) = \frac{kw_0^2}{2} \frac{(1+\zeta^2)}{\zeta} = z_R(\zeta + \zeta^{-1}). \qquad (18)$$

If the beam is not strongly focused, neglecting Gouy's phase in this derivation is a valid approximation, because $|\zeta|$ has large values and arctan($\zeta$) changes only slowly with changing $\zeta$. Thus, the electric field amplitude acquires the compact form

$$E_0 = \frac{1}{w(\zeta)} \exp\left[ ik\left(z + \frac{\rho^2}{2R(\zeta)}\right) - \frac{\rho^2}{w(\zeta)^2} - i\psi_0(\zeta) \right]. \qquad (19)$$

The intensity distribution $|E_0|^2$ of this solution is visualized in Fig.1 for a Gaussian mode with beam waist $w_0 = 4\lambda$. Displayed are sections perpendicular to the propagation axis at equally spaced positions with 250 $\lambda$ distance. Each section is drawn until the intensity $|E_0|^2$ has fallen off to $1/e^5$ of its maximum value at $\rho = 0$. The increasing diameter of the drawn sections is thus a measure of the beam divergence. Fig.1 shows also the phase of the electric field. The occurrence of rings (phase changes) within the planar cross-sections reflects the curvature of the wavefronts, causing the phase to change increasingly faster with increasing $z$ when moving away from the axis $\rho = 0$. The change of the phase along the optical axis from its value 0 to $2\pi$ reflects the „phase jump" of the electric field when propagating through the point of minimum extension (beam waist) as described by Gouy's phase.

Before moving on to the derivation of higher-order Gaussian beams in the next sections, we will shortly stop here for looking at the interaction of a Gaussian beam with an ideal, infinitely thin lens. The result which will be derived here can be easily generalized for the whole family of Gaussian beams the we will consider below. For considering the action of a lens onto a Gaussian beam, it is convenient to introduce the complex curvature $q$ defined by

$$\frac{ik}{2q} = \frac{1-i\zeta}{w^2(\zeta)} \tag{20}$$

i.e.

$$q = \frac{ik}{2}w_0^2(1+i\zeta) = iz_R(1+i\zeta) = iz_R - z, \tag{21}$$

so that the electric field amplitude takes the form

$$E = \frac{1}{w(\zeta)}\exp\left[ik\left(z - \frac{r^2}{2q(\zeta)} - \frac{\psi_0(\zeta)}{k}\right)\right]. \tag{22}$$

Upon interaction with an ideal lens with focal length $f$, a phase shift $-ikr^2/2f$ is added to the phase of the electric field. Remarkably, the thus changed electric field resembles again that of a Gaussian beam, but with transformed complex curvature $q'$ defined by

$$\frac{1}{q'} = \frac{1}{q} + \frac{1}{f}. \tag{23}$$

Of course, the "new" Gaussian beam has to be calculated with respect to a new $z$-axis $z'$ shifted with respect to $z$ so that the new focus position is at $z' = 0$. If $z_L, z'_L$ are the lens positions in the old and new coordinate system, respectively, one then finds

$$\begin{aligned} q' &= iz'_R - z'_L \\ &= \left(\frac{1}{iz_R - z_L} + \frac{1}{f}\right)^{-1} \\ &= \frac{f(ifz_R - fz_L + z_L^2 + z_R^2)}{(z_L - f)^2 + z_R^2} \end{aligned} \tag{24}$$

so that the new Rayleigh length $z'_R$ and lens position $z'_L$ in the new coordinate system are given by the imaginary and real part of the last expression, i.e. by

$$z'_R = \frac{f^2 z_R}{(z_L - f)^2 + z_R^2} \qquad (25)$$

and

$$z'_L = -\frac{f(z_L^2 + z_R^2 - f z_L)}{(z_L - f)^2 + z_R^2}. \qquad (26)$$

As a special example consider the focusing of a Gaussian laser beam when the beam waist is positioned at the focus of the lens, i.e. $z_L = f$. The new waist position is at distance $-z'_L = f$ behind the lens, and the focused beam has the new Rayleigh length $z'_R = f^2/z_R$ corresponding to the new beam waist diameter of $w'_0 = f\lambda/\pi w_0$ being inversely proportional to that of the incident beam.

**Hermite-Gaussian modes**

Hermite-Gaussian beams are a family of structurally stable laser modes which have rectangular symmetry along the propagation axis. In order to derive such modes, the simplest approach is to include an additional modulation of the form $k_x^m k_y^n$ into the amplitude function $A(k_x, k_y)$, with some integer values $n$ and $m$. Then, the electric field amplitude has the form

$$\tilde{E}_{m,n}^H = \int_{-\infty}^{\infty} \frac{dk_x}{2\pi} \int_{-\infty}^{\infty} \frac{dk_y}{2\pi} (ik_x)^m (ik_y)^n e^S \qquad (27)$$

where the abbreviation

$$S(k_x, k_y, x, y, z) = ik_x x + ik_y y + ikz - \frac{w_0^2}{4}(1+i\zeta)(k_x^2 + k_y^2) \tag{28}$$

was used, and the additional factor $i^{n+m}$ was introduced for later convenience. Taking into account the relation

$$ik_x e^S = \frac{\partial}{\partial x} e^S \tag{29}$$

(similarly for the y-direction), $\tilde{E}_{m,n}^H$ can be rewritten in the convenient form

$$\tilde{E}_{m,n}^H = \frac{\partial^{m+n}}{\partial x^m \partial y^n} E_0. \tag{30}$$

Thus, the new field modes occur to be differential derivatives of the fundamental Gaussian mode $E_0$. Looking at the explicit form $E_0$ as given by Eq.(13) shows that the differentiations in the last equation lead to expressions of the form $\partial^p \exp(-\alpha x^2)/\partial x^p$ with some constant $p$ and $\alpha$. Using now the definition of Hermites' polynomials,

$$H_p(x) = (-1)^p \exp(x^2) \frac{d^p}{dx^p} \exp(-x^2), \tag{31}$$

the field amplitude then adopts the form

$$\tilde{E}_{m,n}^H = \frac{1}{w(\zeta)^{(m+n)/2+1}} H_m\left(\frac{x}{w_0(1+i\zeta)^{1/2}}\right) H_n\left(\frac{y}{w_0(1+i\zeta)^{1/2}}\right) \exp\left[ikz - \frac{\rho^2}{w_0^2(1+i\zeta)} - i\tilde{\psi}_{m,n}\right], \tag{32}$$

with the modified Gouy phase

$$\tilde{\psi}_{m,n} = \left(1 + \frac{m+n}{2}\right) \arctan \zeta. \tag{33}$$

The found solution includes complex-valued arguments in both the exponential and pre-exponential functions, and is known as the "elegant" representation of Hermite-Gaussian modes, because of its mathematical elegancy. However, it is more convenient to have complex values only in the exponential function, allowing a clear distinction between the field amplitude and phase. To achieve that one can try to use a modified amplitude modulation in Eq.(27). Instead of using the pure polynomials of $ik_{x,y}$, one uses polynomials of the operator $ik_{x,y} + u^{-1} \partial/\partial k_{x,y}$ with some constant $u$ that has still to be specified. Then, the electric field takes the form

$$E_{m,n}^H \propto \int_{-\infty}^{\infty} \frac{dk_x}{2\pi} \int_{-\infty}^{\infty} \frac{dk_y}{2\pi} \left( ik_x + \frac{1}{u}\frac{\partial}{\partial k_x} \right)^m \left( ik_y + \frac{1}{u}\frac{\partial}{\partial k_y} \right)^n e^S, \qquad (34)$$

This representation also obeys the scalar wave equation because the operators in front of $\exp S$ do not depend on $x$ or $y$ and commute with $\partial/\partial x$ and $\partial/\partial y$. Eq.(34) leads to a new family of solutions for the electric field of a propagating laser beam, parameterized by the arbitrary parameter $u$. Here, we will seek the special value of $u$ that leads to a final result where all pre-exponential functions depend on real arguments only. By using the identity

$$\frac{1}{u}\frac{\partial}{\partial k_x} e^S = \left[ \frac{ix}{u} - \frac{w_0^2}{2u}(1+i\zeta)k_x \right] e^S, \qquad (35)$$

which can be proofed directly by inserting $S$ from Eq.(28), Eq.(34) can be cast into the form

$$E_{m,n}^H = f^{m+n} \left( \frac{\partial}{\partial x} + \frac{ix}{uf} \right)^m \left( \frac{\partial}{\partial y} + \frac{iy}{uf} \right)^n E_0. \qquad (36)$$

where the abbreviation

$$f = 1 - \frac{w_0^2}{2iu}(1+i\zeta) \tag{37}$$

is introduced.

Next, the important operator identity

$$\left(\frac{\partial}{\partial x} + \alpha x\right)^p \equiv \left[\exp\left(-\frac{\alpha x^2}{2}\right)\frac{\partial}{\partial x}\exp\left(\frac{\alpha x^2}{2}\right)\right]^p = \exp\left(-\frac{\alpha x^2}{2}\right)\frac{\partial^p}{\partial x^p}\exp\left(\frac{\alpha x^2}{2}\right) \tag{38}$$

will be exploited which can be checked directly by explicitly performing the differentiation within the square bracket. This operator identity will be used in this paper several times for simplifying expression involving terms like $(\partial/\partial x + \alpha x)^p$ with constant $p$ and $\alpha$. Indeed, using the identity of Eq.(38) allows us to rewrite Eq.(36) as

$$\begin{aligned} E_{m,n}^H &\propto f^{m+n}\left[\exp\left(-\frac{ix^2}{2uf}\right)\frac{\partial}{\partial x}\exp\left(\frac{ix^2}{2uf}\right)\right]^m \left[\exp\left(-\frac{iy^2}{2uf}\right)\frac{\partial}{\partial y}\exp\left(\frac{iy^2}{2uf}\right)\right]^n E_0 \\ &= f^{m+n}\exp\left(-\frac{i\rho^2}{2uf}\right)\frac{\partial^m}{\partial x^m}\frac{\partial^n}{\partial y^n}\exp\left(\frac{i\rho^2}{2uf}\right)E_0. \end{aligned} \tag{39}$$

Going again back to the explicit expression for $E_0$, Eq.(13), it is seen that the differential operators in the last equation now act on an exponential function containing the argument

$$\frac{i\rho^2}{2uf} - \frac{\rho^2}{w_0^2(1+i\zeta)} = -\frac{\rho^2}{2iu - w_0^2(1+i\zeta)} - \frac{\rho^2}{w_0^2(1+i\zeta)} \tag{40}$$

where, on the r.h.s., the explicit form of $f$ as given by Eq.(37) was used. Obviously, this expression becomes real if the first summand is complex conjugated to the second one, i.e. if $u = -iw_0^2$, leading to

$$f = 1 - \frac{w_0^2}{2iu}(1+i\zeta) = \frac{1-i\zeta}{2} \tag{41}$$

which is exactly what we wanted to achieve: the exponential function on which the differential operators are acting are real valued functions, so that the differentiations lead to real valued pre-exponential functions. Indeed, using the last two equations, the electric field amplitude is proportional to

$$\begin{aligned} E_{m,n}^H &\propto f^{m+n} \exp\left(-\frac{ir^2}{2uf}\right) \frac{\partial^m}{\partial x^m} \frac{\partial^n}{\partial y^n} \exp\left(\frac{i\rho^2}{2uf}\right) E_0 \\ &= \frac{f^{m+n}}{\pi w_0 w} \exp\left(-\frac{ir^2}{2uf}\right) \frac{\partial^m}{\partial x^m} \frac{\partial^n}{\partial y^n} \exp\left[ikz - \frac{2\rho^2}{w_0^2(1+\zeta^2)} - i\psi_0\right] \\ &= \frac{(-f)^{m+n}}{\pi w_0 w} \left(\frac{2}{w_0^2(1+\zeta^2)}\right)^{\frac{m+n}{2}} H_m\left(\sqrt{2}\frac{x}{w}\right) H_n\left(\sqrt{2}\frac{y}{w}\right) \exp\left[ikz - \frac{\rho^2}{w_0^2(1+i\zeta)} - i\psi_0\right] \\ &= \frac{(-1)^{m+n}}{2^{(m+n)/2} \pi w_0^{m+n+1} w} H_m\left(\sqrt{2}\frac{x}{w}\right) H_n\left(\sqrt{2}\frac{y}{w}\right) \exp\left[ikz - \frac{\rho^2}{w_0^2(1+i\zeta)} - i\psi_{m,n}\right] \end{aligned} \tag{42}$$

where the new Gouy phase is defined by $\psi_{m,n} = (1+m+n)\arctan\zeta$. After dropping all constant factors in this expression, one arrives at the standard form of the *Hermite-Gauss* mode:

$$E_{m,n}^H = \frac{1}{w(\zeta)} H_m\left(\sqrt{2}\frac{x}{w(\zeta)}\right) H_n\left(\sqrt{2}\frac{y}{w(\zeta)}\right) \exp\left[ikz - \frac{\rho^2}{w_0^2(1+i\zeta)} - i\psi_{m,n}\right]. \tag{43}$$

For visualizing of what we have obtained, Fig.2 shows the real part of the amplitudes within the plane $z=0$ for different values of *m, n* in the range of $m+n=0\ldots3$. The top panel shows the profile of the fundamental Gaussian mode, $E_{0,0}^H \equiv E_0$. From the figure, as well as from the method of construction of the Hermite-Gauss modes as derivatives of the fundamental mode $E_0$, one sees that the total number of occurring maxima and minima is given by the sum $m+n+1$, and that one has the symmetry relation $E^H_{m,n}(x,y,z) = E^H_{n,m}(y,x,z)$.

**Laguerre-Gauss modes**

Differently from Hermite-Gaussian beams, Laguerre-Gaussian modes have rotational symmetry along their propagation axis and carry an intrinsic rotational orbital angular momentum of $i\hbar$ per photon [10, 11 and references therein]. This means that a refractive object placed along the propagation axis will experience a torque. This property of Laguerre-Gaussian beams is of considerable practical interest, particularly in the field of optical trapping and for driving micromachined elements with light. It is important to point out that this intrinsic rotational momentum has to be distinguished from the angular momentum due to the polarization of light. Laguerre-Gaussian modes can be derived by modulating the amplitude function $A(k_x, k_y)$ with a periodic function of the angular variable $\alpha = \arctan(k_y/k_x)$. Taking into account the identity

$$k_x \pm ik_y = \sqrt{k_x^2 + k_y^2}\, e^{i\alpha}, \tag{44}$$

such a modulation is equivalent to using an amplitude function proportional to $(k_x + ik_y)^m (k_x - ik_y)^{n+m}$ with some integer values $m$ and $n$, resulting in an electric field with $\exp(-in\phi)$ dependency on the angular variable $\phi = \arctan(y/x)$. Now, the electric field acquires the form

$$\tilde{E}^L_{m,n} \propto \int_{-\infty}^{\infty} \frac{dk_x}{2\pi} \int_{-\infty}^{\infty} \frac{dk_y}{2\pi} (k_x + ik_y)^m (k_x - ik_y)^{n+m} e^S. \tag{45}$$

Using the same ideas as in the previous section, this leads to

$$\tilde{E}^L_{m,n} \propto \left(\partial_x + i\partial_y\right)^m \left(\partial_x - i\partial_y\right)^{n+m} E_0 \tag{46}$$

or, by using the substitutions $\omega = x + iy$, $\overline{\omega} = x - iy$, $\partial_x + i\partial_y = 2\partial_{\overline{\omega}}$, and $\partial_x - i\partial_y = 2\partial_\omega$ to

$$\tilde{E}^L_{m,n} \propto \partial_{\overline{\omega}}^m \partial_\omega^{n+m} E_0. \tag{47}$$

When taking into account that $\rho^2 = x^2 + y^2 = \omega\overline{\omega}$, and employing the definition of Laguerre's functions $L^n_m$,

$$L^n_m(r) = \frac{e^r r^{-n}}{m!} \frac{d^m}{dr^m}\left(e^{-r} r^{n+m}\right), \tag{48}$$

the differentiation of the fundamental mode leads to

$$\begin{aligned}\partial_{\overline{\omega}}^m \partial_\omega^{n+m} E_0 &= \partial_{\overline{\omega}}^m \left(-\frac{\overline{\omega}}{w_0^2(1+i\zeta)}\right)^{n+m} E_0 \\ &= (-\omega)^{-n-m} m! \left(\frac{\omega\overline{\omega}}{w_0^2(1+i\zeta)}\right)^n \left(\frac{\omega}{w_0^2(1+i\zeta)}\right)^m L^n_m\left(\frac{\omega\overline{\omega}}{w_0^2(1+i\zeta)}\right) E_0.\end{aligned} \tag{49}$$

After omitting any constant factors one finally has

$$\tilde{E}^L_{m,n} = \frac{e^{-in\phi}}{w(\zeta)^{n+m+1}} \rho^n L^n_m\left(\frac{\rho^2}{w_0^2(1+i\zeta)}\right) \exp\left[ikz - \frac{\rho^2}{w_0^2(1+i\zeta)} - i\tilde{\psi}^L_{m,n}\right], \tag{50}$$

with the new phase $\tilde{\psi}^L_{m,n} = (n+m+1)\arctan\zeta$.

As in the previous section, this mode displays complex valued arguments in both the exponential and pre-exponential functions. To rectify this situation, one can again apply the same idea as in the previous section and modify the amplitude modulation to

$$E_{m,n}^L \propto \int_{-\infty}^{\infty} \frac{dk_x}{2\pi} \int_{-\infty}^{\infty} \frac{dk_y}{2\pi} \left[ i(k_x + ik_y) + \frac{1}{u}\left(\frac{\partial}{\partial k_x} + i\frac{\partial}{\partial k_y}\right) \right]^m \left[ i(k_x - ik_y) + \frac{1}{u}\left(\frac{\partial}{\partial k_x} - i\frac{\partial}{\partial k_y}\right) \right]^{n+m} e^S \quad (51)$$

with the same value of $u$ as before, $u = -iw_0^2$. This leads directly to

$$\begin{aligned}
E_{m,n}^L &\propto f^{2m+n} \left( 2\partial_{\bar{\omega}} + \frac{i\omega}{uf} \right)^m \left( 2\partial_\omega + \frac{i\bar{\omega}}{uf} \right)^{n+m} E_0 \\
&= (2f)^{2m+n} \exp\left(-\frac{i\omega\bar{\omega}}{2uf}\right) \partial_{\bar{\omega}}^m \left(-\frac{2\bar{\omega}}{w^2}\right)^{n+m} \exp\left(\frac{i\omega\bar{\omega}}{2uf}\right) E_0 \\
&= (2f)^{2m+n} (-\omega)^{-n-m} m! \left(\frac{2\omega\bar{\omega}}{w^2}\right)^n \left(\frac{2\omega}{w^2}\right)^m L_m^n\left(\frac{2\omega\bar{\omega}}{w^2}\right) E_0 \\
&= \frac{(-1)^{n+m} 2^{m+n} m! e^{-in\phi}}{w^{2(n+m)}} \rho^n L_m^n\left(\frac{2\rho^2}{w^2}\right) \exp\left[ ikz - \frac{\rho^2}{w_0^2(1+i\zeta)} - i\psi_{m,n}^L \right]
\end{aligned} \quad (52)$$

where the same definition of $f$ was used as in the previous section, and an identity similar to Eq.(38) was exploited. In the last line, the new phase $\psi_{n,m}^L = (n + 2m + 1)\arctan\zeta$ was defined. After omitting all constant factors, the standard definition of a *Laguerre-Gauss* mode is obtained:

$$E_{m,n}^L = \frac{e^{-in\phi}}{w(\zeta)} \left(\frac{\rho}{w(\zeta)}\right)^n L_m^n\left(\frac{2\rho^2}{w^2(\zeta)}\right) \exp\left[ ikz - \frac{\rho^2}{w_0^2(1+i\zeta)} - i\psi_{n,m}^L \right]. \quad (53)$$

Similarly to the visualization of the Hermite-Gauss modes in the previous section, Fig.3 shows the real part of the obtained amplitudes within the plane $z = 0$ up to $n, m = 3$. As can be seen, the real part of the amplitudes are rotationally symmetric around the axis $z$, and the number $m + 1$ is the number of minima and maxima along a radial line from the center $\rho = 0$ (note that the first minimum always occurs at $\rho = 0$, where the amplitude vanishes). The number $n$ defines by how many full cycles of $2\pi$ the phase is changing when one goes around the axis of propagation $\rho = 0$, and thus defines the *chirality* of the mode. This phase behavior for $n = 1\ldots3$ is shown in Fig.4, where surfaces of constant phase, $\arg E_{m,n}^L \equiv 0 \,(\mathrm{mod}\, 2\pi)$, near the axis $\rho = 0$ are shown.

**Bessel beams**

Any light beam whose energy is confined to a finite region around its axis of propagation is subject to diffractive spreading as it propagates in free space. The Rayleigh length is the characteristic distance beyond which diffraction becomes increasingly noticeable. This is the case for all beams that we have analyzed so far: Gaussian, Hermite-Gaussian or Laguerre-Gaussian beams with beam waist $w_0$ diverging with a divergence angle $\arctan\left(\lim_{\zeta\to\infty} w(\zeta)/z\right) = \arctan(\lambda/\pi w_0)$. However, there exists a family of laser beams which maintain a constant intensity profile along the propagation direction, so-called *non-diffracting beams*, which were first described by Durnin in 1987 [16]. They can be easily derived within the framework of the plane-wave representation Eq.(1) by choosing the special, δ-function modulation function $A(k_x, k_y) \sim \delta\left(\sqrt{k_1^2 + k_2^2} - k_\perp\right)$ with some arbitrary parameter $0 < k_\perp < k = 2\pi\lambda$. The resulting electric field amplitude distribution then reads

$$E \propto \exp\left(i\sqrt{k^2 - k_\perp^2}\, z\right) \int_0^{2\pi} \frac{d\psi}{2\pi} \exp(ik_\perp \rho \cos\psi) = \exp\left(i\sqrt{k^2 - k_\perp^2}\, z\right) J_0(k_\perp \rho) \tag{54}$$

where $J_0$ denotes the zero-order Bessel function of the first kind, which lends this family of beams also the name *Bessel beams*. When $k_\perp = 0$, one obtains simply a plane wave, but for positive values of $k_\perp$ the solution is a non-diffracting beam whose intensity profile decays inversely proportional to $k_\perp \rho$. The width of the central intensity maximum is determined by $k_\perp$, reaching a minimum value of approximately $3\lambda/4$ when $k_\perp = k = 2\pi\lambda$. Obviously, the intensity distribution $|E|^2$ of Bessel beams does no longer depend on the variable *z*. Bessel beams resemble narrow light rods with a surrounding concentric ring structure, propagating through space without divergence. In sharp contrast to Gaussian beams, the energy density of Bessel beams is no longer confined to the vicinity of the propagation axis: The integral

$$\langle \rho^2 \rangle = \int_0^\infty d\rho \rho^3 |E|^2 \propto \int_0^\infty d\rho \rho^3 J_0^2(k_\perp \rho)$$

is highly divergent. Thus, an ideal Bessel beams cannot be realized in practice, since producing a beam whose transversal profile is invariant with longitudinal distance would require an infinite aperture and infinite energy. However, approximations of Bessel beams can be created experimentally, for example by using an axicon (a conical prism) [20, 21], an annular aperture, or diffractive optical elements [23]. An important application of approximate Bessel beams has emerged in special setups for optical trapping. Since the diffraction-limited region of an approximate Bessel beam extends over large propagation distances, multiple trapping of particles with a single beam becomes possible [19]. A further important characteristic of non-diffracting beams is their ability to self-reconstruction after interaction with an obstacle [18]: Even if the central intensity maximum of a Bessel beam is completely blocked by an object, the beam reforms its original field distribution after some propagation distance. Several practical applications [20, 23] of this property have been suggested.

**Conclusion**

The plane wave representation of Eq.(1), that was the starting point for deriving the different beam modes, has the significant advantage that it is easy to generalize. Other cases of interest can be readily analyzed. For example, it is straightforward to derive beam modes with non-axial symmetry by using an amplitude function of the form $A(k_x, k_y) \propto \exp[-(w_{0x}^2 k_x^2 + w_{0y}^2 k_y^2)/4]$, resulting in elliptical beams with the two principal planes *xz* and *yz*, and a set of two beam parameters $w_{0x,y}$. Such beams are also called *astigmatic*, because generally the position of the beam waist along the axis of propagation will be different within the two principal planes. For several applications, e.g. when considering the propagation of laser light in optically anisotropic media, it is necessary to take into account the vector character of the electromagnetic field [4]. This can be easily done by replacing the scalar amplitude function in Eq.(1) by a vector function which has to be chosen in such a way that the direction of the amplitude vector is perpendicular to the wave vector of the corresponding plane wave component, ensuring that the whole representation automatically obeys the vector wave equation. For example, the vector amplitude function of an *x*-polarized beam has then the form $\mathbf{A}(k_x, k_y) \propto (\hat{\mathbf{k}} \times \hat{\mathbf{y}}) A(k_x, k_y)$, where $A(k_x, k_y)$ is similar to amplitude functions as used in the preceding sections. Finally, for tightly focused beams, the paraxial condition does no longer hold (see e.g. [24]), so that the extension of the integrals in Eq.(1) to infinity and the expansion of the root $(k^2 - k_x^2 - k_y^2)^{1/2} \approx k - (k_x^2 + k_y^2)/2k$ is no longer justified. Nonetheless, the representation in Eq.(1) (and its vector analog) still holds and is a perfect starting point for numerical evaluation of non-paraxial beams.

**Figures**

Fig.1: Intensity distribution $|E_0|^2$ (left panel) and phase $\arg E_0$ (right panel) of the fundamental Gaussian modes across planes perpendicular to the axis $\rho = 0$, spaced at distances of $250\lambda$ and starting at $z = -1000\lambda$. The beam waist diameter was set to $w_0 = 4\lambda$. Each section is drawn until the intensity $|E_0|^2$ has fallen off to $1/e^5$ of its maximum value at $\rho = 0$. All axis units are in wavelengths. At each cross-section, intensity values are normalized by heir on-axis maximum value.

Fig.2: Real part of the amplitude of Hermite-Gauss modes, $\operatorname{Re}(E_{m,n}^H)$, within the plane $z = 0$, for laser beams with beam waist radius $w_0 = 4\lambda$. Shown are square areas of size $20\lambda \times 20\lambda$. Amplitudes are normalized to their maximum value. Number in curly brackets refer to the mode numbers $m$ and $n$.

Fig.3: Real part of the amplitude of Laguerre-Gauss modes, $\operatorname{Re}(E_{m,n}^L)$, within the plane $z = 0$, for laser beams with beam waist radius $w_0 = 4\lambda$. Shown are square areas of size $20\lambda \times 20\lambda$. Amplitudes are normalized to their maximum value. Number in curly brackets refer to the mode numbers $m$ and $n$.

Fig.4: Chiral phase behavior of Laguerre-Gauss modes. Shown are surfaces of constant phase $\arg E_{m,n}^L \equiv 0 \,(\mathrm{mod}\, 2\pi)$ for (a) $n = 1$, (b) $n = 2$, and (c) $n = 3$. The units of all three axes are in wavelengths, the vertical axis is the axis of propagation ($\rho = 0$). The number $n$ not only determines the swiftness of phase change when moving around the vertical axis, but also the number of non-connected surfaces with the same phase value.

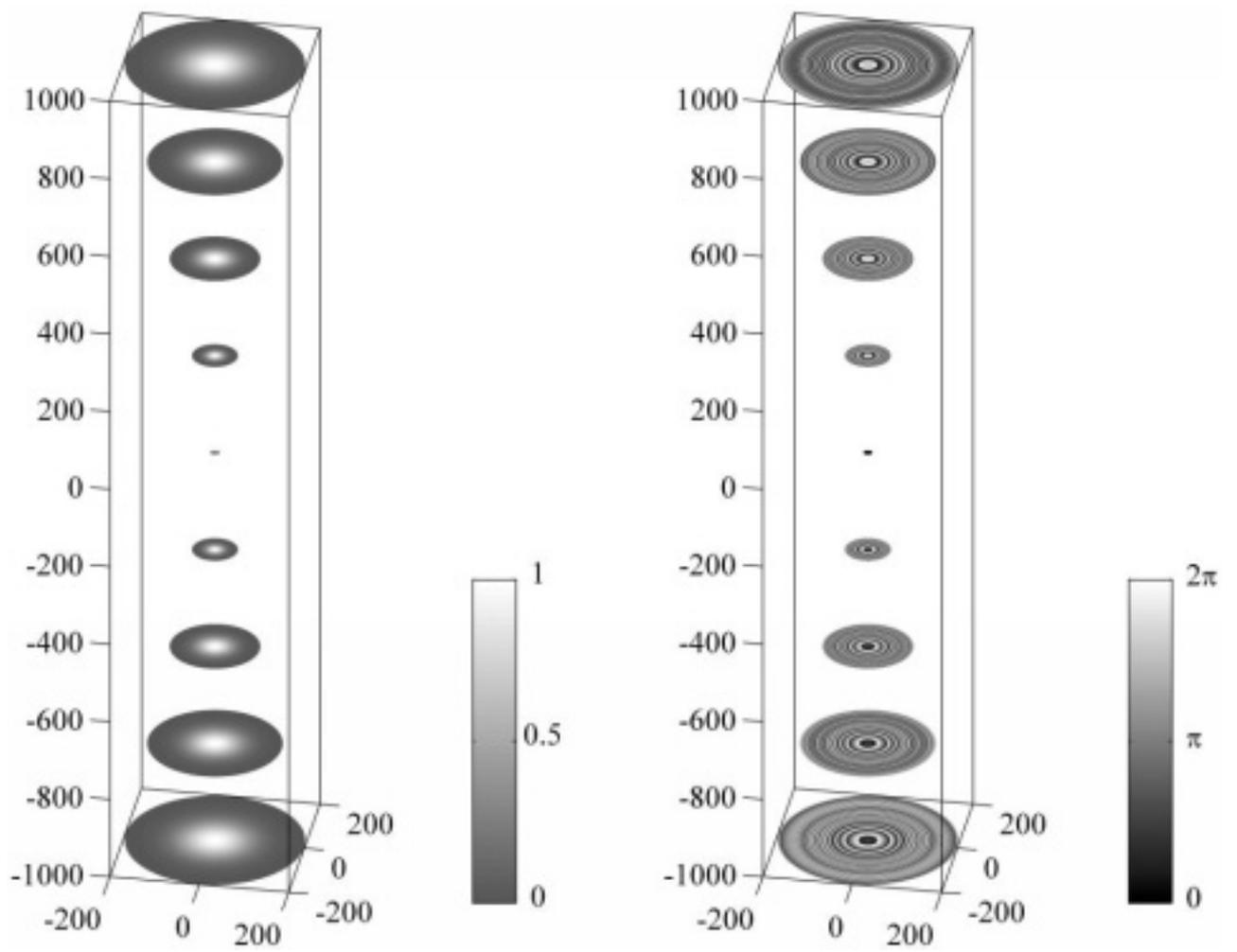

Fig. 1

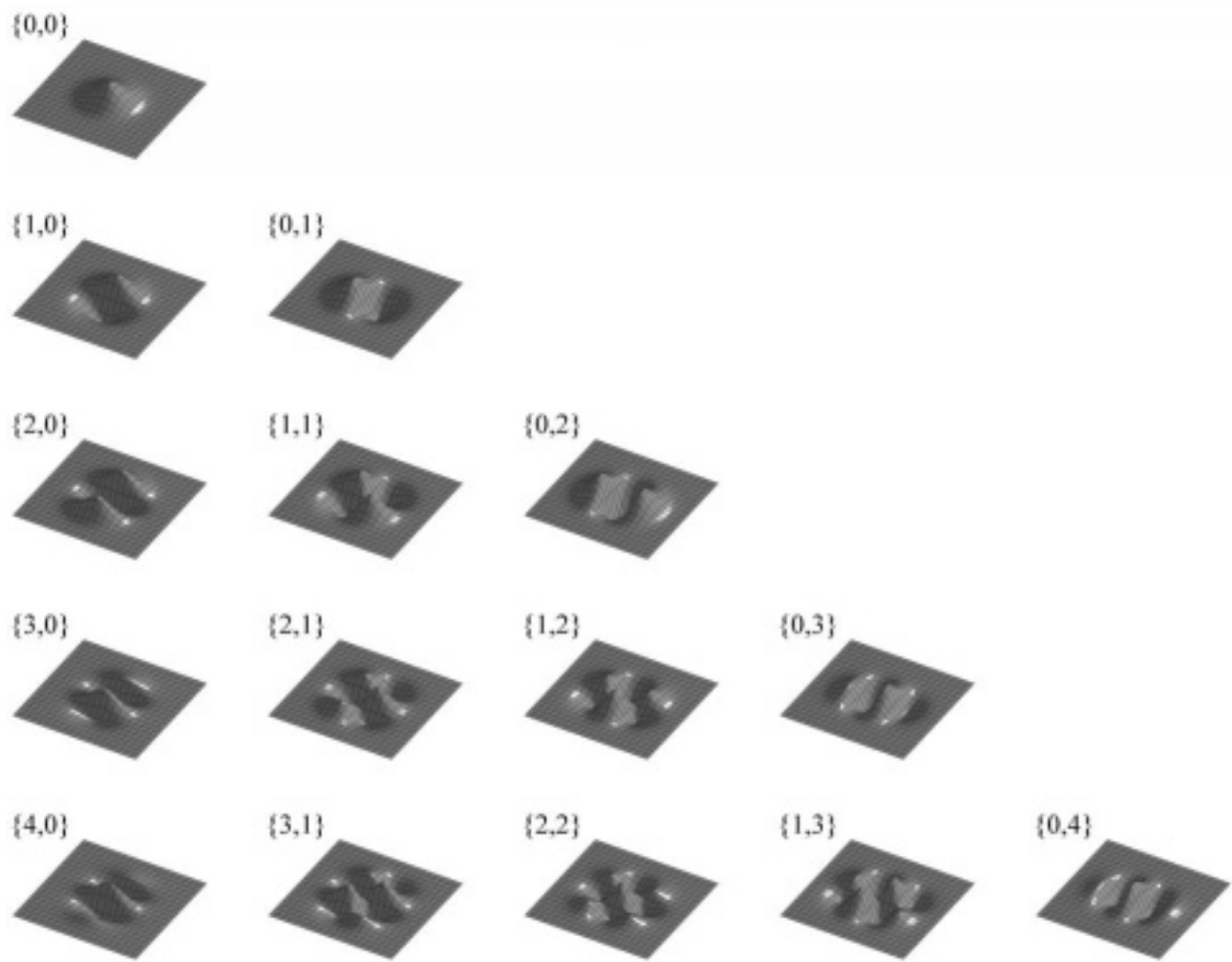

Fig. 2

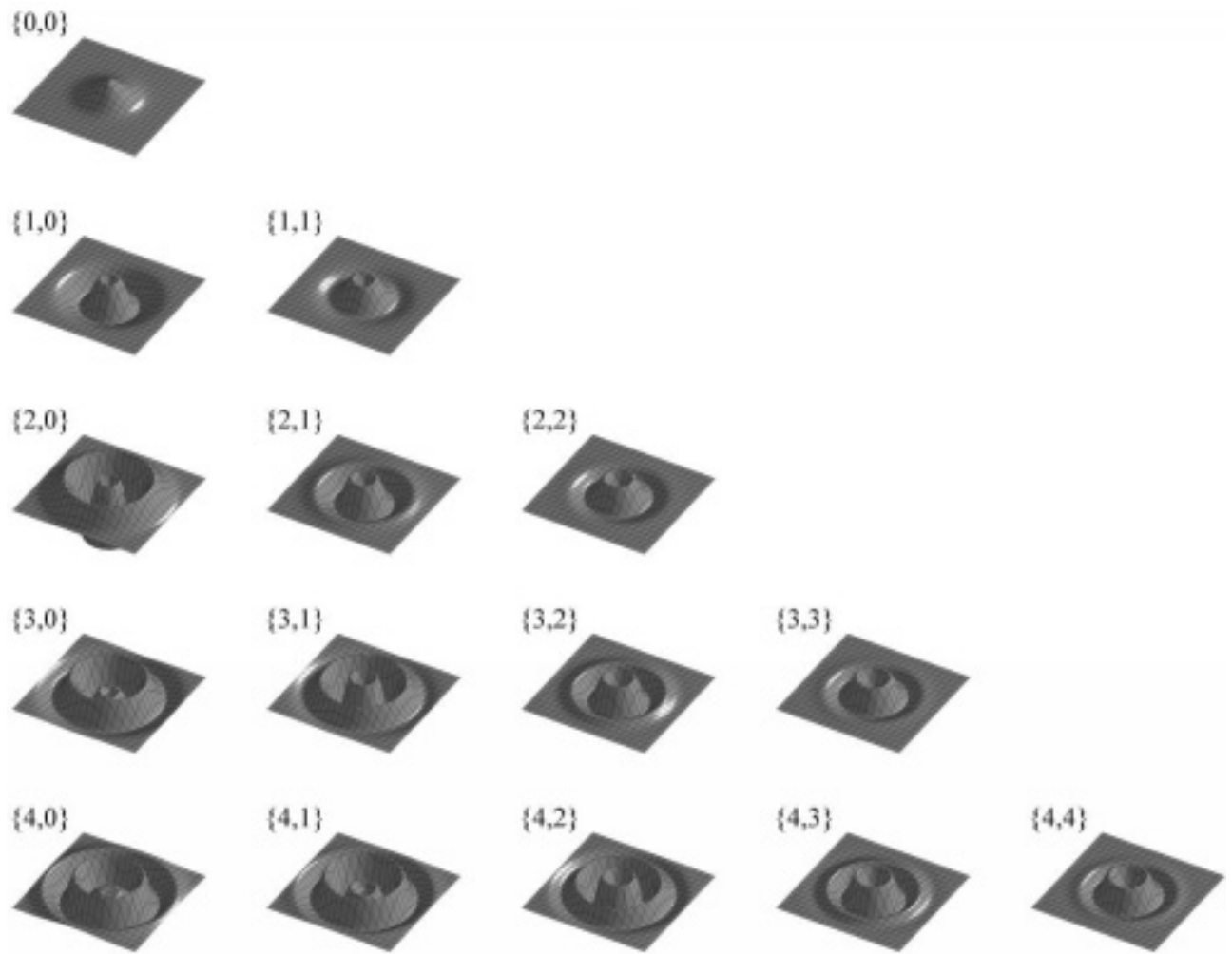

Fig. 3

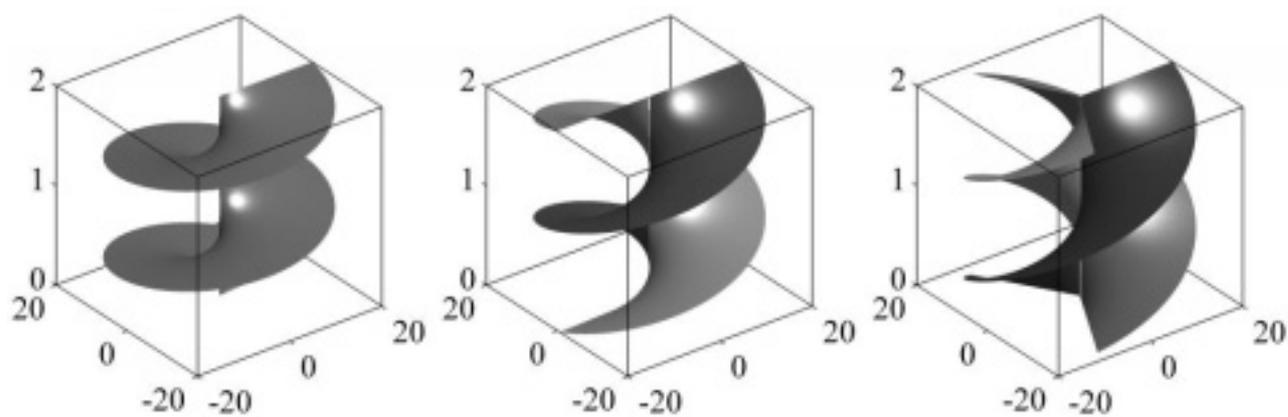

Fig. 4